\documentclass[12pt,preprint]{aastex}

\newcommand{\PA}{POINT-AGAPE }
\newcommand{\PAcol}{POINT-AGAPE collaboration }
\newcommand{\ps}{power spectrum }
\newcommand{\psa}{power spectra }
\newcommand{\psns}{power spectrum}
\newcommand{\psans}{power spectra}
\newcommand{\nh}{$N_{H}$ }
\newcommand{\nhns}{$N_{H}$}
\newcommand{\prob}{probability }

\newcommand{\probs}{probabilities }
\newcommand{\probsns}{probabilities}
\newcommand{\lcns}{lightcurve}
\newcommand{\lc}{lightcurve }
\newcommand{\lcsns}{lightcurves}
\newcommand{\lcs}{lightcurves }
\newcommand{\kfold}{$K$-fold cross-validation }
\newcommand{\kfoldns}{$K$-fold cross-validation}
\newcommand{\rp}{RpropMAP }
\newcommand{\rpns}{RpropMAP}

\slugcomment{To appear in The Astronomical Journal}

\shorttitle{Detection of Classical Novae with Neural Networks}
\shortauthors{Belokurov et al.}

\begin{document}


\title{Automated Detection of Classical Novae with Neural Networks}


\author{S.M. Feeney, V. Belokurov, N.W. Evans, J. An, P.C. Hewett}
\affil{Institute of Astronomy, Madingley Rd, Cambridge CB3 0HA, UK}

\author{M. Bode, M. Darnley, E. Kerins}
\affil{Astrophysics Research Institute, Liverpool John Moores University,
Twelve Quays House, Egerton Wharf, Birkenhead CH41 1LD, UK}

\author{P. Baillon}
\affil{European Organization for Nuclear Research CERN,
CH-1211 Gen\`eve 23, Switzerland}

\author{B.J. Carr}
\affil{Astronomy Unit, School of Mathematical Sciences, Queen Mary,
University of London, Mile End Road, London E1 4NS, UK}

\author{S.~Paulin-Henriksson}
\affil{Laboratoire de Physique Corpusculaire et Cosmologie, Coll\`ege de
France, 11 Place Marcelin Berthelot, F-75231 Paris, France}

\author{A. Gould}
\affil{Department of Astronomy, Ohio State University,
140 West 18th Avenue, Columbus, OH 43210, USA}


\begin{abstract}
The \PAcol surveyed M31 with the primary goal of optical detection of
microlensing events, yet its data catalogue is also a prime source of
\lcs of variable and transient objects, including classical novae
(CNe). A reliable means of identification, combined with a thorough
survey of the variable objects in M31, provides an excellent
opportunity to locate and study an entire galactic population of
CNe. This paper presents a set of 440 neural networks, working in 44
committees, designed specifically to identify fast CNe. The networks
are developed using training sets consisting of simulated novae and
\PA \lcs in a novel variation on \kfoldns, and use the binned,
normalised \psa of the \lcs as input units. The networks successfully
identify 9 of the 13 previously identified M31 CNe within their
optimal working range (and 11 out of 13 if the network error bars are
taken into account). The networks provide a catalogue of 19 new
candidate fast CNe, of which 4 are strongly favoured.
\end{abstract}





\section{Introduction}

One of the greatest advances of modern experimental astrophysics is
the automation of photometric surveys, which allow massive amounts of
data to be gathered systematically, efficiently and with the minimum
need for human intervention. Such surveys scour large regions of the
sky, carefully searching for a wide variety of rare objects and
phenomena such as microlensing events (surveys like OGLE, MACHO and
EROS), gamma-ray burst optical counterparts (ROTSE), extra-solar
planetary transits (SuperWASP) and near-Earth objects (NEAT). These
surveys have provided the scientific community with invaluable
information and resulted in many new discoveries, yet they have also
left us with a new (and very welcome) problem: how can we sort through
the vast data catalogues to reliably filter out objects of interest?

The raw data produced by these surveys are simply collections of the
\lcs of the objects found in the survey's field of detection.
Transient objects hold particular interest for a long list of fields,
including cosmology (SNe Ia), single and binary stellar evolution (SNe
and cataclysmic variables, respectively), and dark matter studies
(microlensing). They are generally rare and have short lifetimes, so
must be identified and studied quickly. The sheer size of such
datasets means that such transient objects are inevitably present in
the catalogues; however there is still a pressing need to detect
objects swiftly and reliably for further study or follow-up.  A number
of researchers have argued that neural networks may provide a viable
solution to this problem \citep{wozniak,vas,vas1,brett}. Neural
networks have already been proven to be useful pattern-recognition
tools in astrophysical applications such as galaxy \citep{lah} and
stellar spectra \citep{bail} classification. They are highly
adaptable, easy and quick to use, but perhaps their most relevant
asset in this application is their ability to attach a probability to
their classification of an object, thus allowing the user to
prioritise their further study.

The contribution of this paper is to provide working neural networks
for the detection of classical novae (CNe). These are close
interacting binary stars, consisting of a white dwarf primary and a
cool red dwarf secondary. The secondary star overflows its Roche lobe
and loses mass to the primary. Very occasionally, runaway
thermonuclear burning of the degenerate layer of hydrogen accreted by
the white dwarf can cause a nova outburst. The nova's brightness rises
rapidly to an absolute magnitude of between $-6$ and $-9$ before
slowly fading back to quiescence. Much remains unknown concerning the
abundance and distribution of nova in galaxies due to the lack of
systematic surveys. So, there is a need for fully automated, and less
subjective, selection of candidate CNe so that more soundly based
conclusions concerning the nova rate and distributions can be
drawn. \cite{darn} have already devised one possible systematic
algorithm. Here, we provide an alternative to the method of Darnley et
al. using a novel application of neural networks.

The paper is organised as follows. In \S 2, the dataset through which
we search for CNe lightcurves is described. This is derived from the
POINT-AGAPE microlensing experiment towards M31. Although the primary
aim of this experiment is to find microlensing events, the dataset of
varying lightcurves is a rich resource for the study of variable stars
towards M31 \citep{jin}. \S 3 discusses the properties of nova
lightcurves and summarises previous work to find CNe in M31. Next, \S
4 provides a short introduction to neural networks for the
astronomical user. \S 5 describes the pre-processing and the
architecture of neural networks to identify CNe, while \S 6 describes
the computations. The nova catalogue obtained by the networks is
presented in \S 7.

\section{The Lightcurve Data Set}

\label{sub:PA}

The data used in this paper was gathered by the \PAcol working with
the Wide Field Camera (WFC) mounted on the 2.5m Isaac Newton telescope
(INT) on La Palma. The collaboration took images of the Andromeda
Galaxy (M31) over the course of three observing seasons (1999-2001),
searching for evidence of microlensing events
\citep{michel,ph1,ph2,beletal}. For one hour of each observing night,
the WFC was used to take images of M31 over two fields, to the north
and south of M31's central bulge, with each field-image formed using
the four 4100$\times$2048 CCDs that make up the WFC (see Figure 1 of
\citet{jin}). The raw data produced by the \PAcol then consisted of
light curves generated from the flux gathered in three pass-bands by
individual pixels in each field-image. The pass-bands used were
denoted \textit{g}, \textit{r} and \textit{i}, and are similar to
those used by the Sloan Digital Sky Survey.  The M31 fields are mainly
composed of unresolved stars, and the effects of seeing from epoch to
epoch are substantial.  In order to build lightcurves, we use the
superpixel method to ensure the same fraction of flux falls within the
window function, irrespective of seeing \citep{annelaure,reza,yann}.
This provides superpixel \lcs (7$\times$7 pixels in size).  Each pixel
is $0\farcs33$ on a side, so the $7 \times 7$ superpixel is
$2\farcs1$ on a side. This matches the typically worst seeing at the
INT site, which is about $2''$. The superpixel \lcs are then
cleaned (for details, see \citet{irw} and \citet{jin}): a mask of the
known CCD defects was constructed, together with regions around all
resolved stars detected in the reference frame. After masking, 44635
variable superpixel $r$ band \lcs remained, and this is the catalogue
through which we search for nova-like \lcsns.

\begin{figure}[tb]
\includegraphics[height=6.5cm]{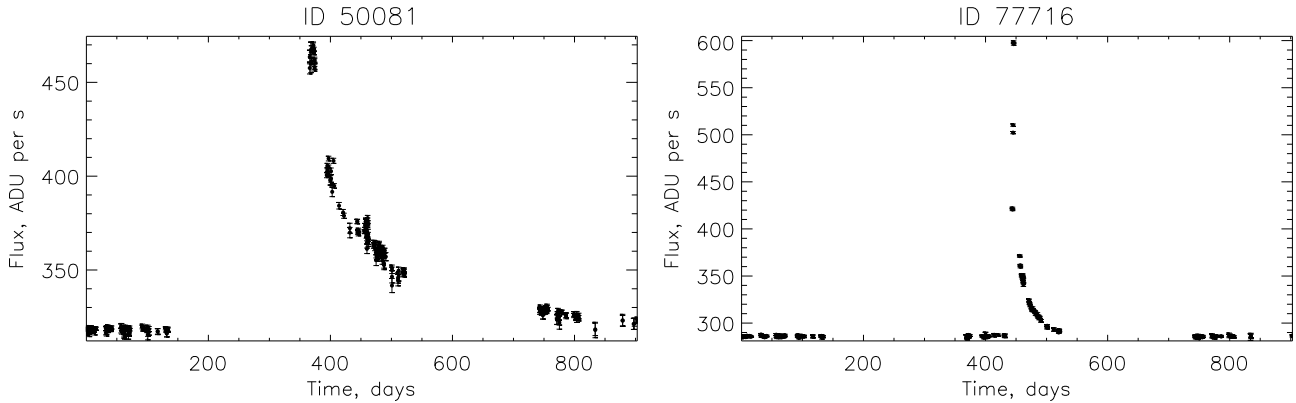}
\caption{Left: Lightcurve of a slow nova in M31, as identified by
Darnley et al. (ID: PACN-00-02). Note the decay fluctuations in the
declining part of the lightcurve. Right: Lightcurve of a fast nova in
M31, as identified by Darnley et al. (ID: PACN-00-06) and An et
al. (ID: 77716).  } \label{M31nova}
\end{figure}

Although the collaboration produced a very large amount of data and
thus greatly increased the chances of discovering new objects, there
are two complicating factors which slightly reduce the data's quality
and ease of analysis. First, the observations were carried out over
the course of three seasons. These seasons correspond to the periods
in which M31 was visible from the northern hemisphere, and mean that
the \lcs are sampled in runs of $\sim$150 days, with $\sim$200-day
gaps (see Figure \ref{M31nova} for an illustration of the
sampling). Three other factors, the limited mounting of the WFC, the
limited scheduled observing time on the INT and the weather, result in
the sampled runs consisting of well-sampled periods typically lasting
1-2 weeks, separated by very poorly-sampled periods lasting 1-3
weeks. Secondly, the large distance of M31 means that in most cases
single stars are not resolved by the INT. This means that the
superpixel \lcs almost always consist of flux produced by more than
one star, which could result in very exotic \lcsns, hence limiting our
ability to classify objects.

\section{Classical Novae in M31}
\label{sub:LCs}

In classical novae, the cool red dwarf secondary overflows its Roche
lobe and loses mass to the primary white dwarf. This mass builds up in
an accretion disc before falling onto the surface of the white dwarf
(see e.g., Bode \& Evans 1989). The main feature of nova \lcs is a
single outburst\footnote{CNe are \emph{required} to have had only one
major outburst in historic times. A few CNe in quiescence show smaller
outbursts, similar to those in dwarf novae, caused by changes in mass
flux through the accretion disc}, typically increasing the absolute
magnitude of the nova to between -6 and -9 before slowly (compared to
the initial rise) fading back to the quiescent state. These classical
nova outbursts are caused by the runaway thermonuclear burning of the
degenerate layer of hydrogen accreted by the white dwarf. Once a
critical amount of hydrogen has been accreted, it begins to burn via
the CNO cycle, precipitating thermonuclear runaway and resulting in
the ejection of the accreted layer on the white dwarf surface. This
explosion and ejection are accompanied by an intense brightening,
followed by a gradual decay back to quiescence.

The progress of the nova outburst depends on several parameters,
including the mass accretion rate from the secondary, and the
temperature and mass of the white dwarf (e.g., Prialnik \& Kovetz
1995). The outbursts therefore vary from system to system, as shown by
the rich viety of CNe lightcurves in Sterken \& Jaschek
(1996). However, it is possible to divide novae into speed classes
according to the time ($t_2$) taken to decline by two magnitudes from
maximum light, the two main classes being fast ($t_2 < 80$ days) and
slow ($t_2 > 80$ days) novae (e.g., Payne-Gaposchkin 1957).  Fast
novae rise rapidly to maximum light, taking 1 to 2 days, and generally
have relatively smooth initial decays with only small fluctuations in
their early light curves. Slow novae on the other hand can take much
longer to reach maximum light and usually have more erratic lightcurve
decays, with strong fluctuations capable of producing secondary maxima
of varying strengths during initial decline.  Furthermore, the maximum
absolute magnitudes of classical novae are correlated to the rate of
their decline, which coupled with their high luminosities makes
classical novae potentially important standard candles \citep{hubble,cohen}.
Figure \ref{M31nova} shows the \lcs of a slow and fast nova
respectively, as previously found in M31.

The \lcs of classical novae share many features with the \lc peaks of
dwarf novae and recurrent novae. The main distinguishing feature in
the \lcs of these objects is that dwarf and recurrent novae undergo
repeated outbursts. However, the periods between outbursts and the
gaps in the \PA sampling could lead to only one peak of a dwarf or
recurrent nova \lc being sampled. Hence, we may pick up some stray
dwarf or recurrent novae in our final catalogue. Dwarf nova outbursts
are not be detectable in M31. However, they may be present in the \PA
catalogue as foreground objects, though even this has a very low
probability.

Dedicated nova searches of M31 have been carried out ever since Hubble
first did so in 1929 (see Table 1 of \citet{darn} for a list of
papers). Very recently, \citet{darn} and \citet{jin} have published
CNe lightcurves from the \PA catalogue. \citet{darn} used a pipeline
(see Table 3 in their paper) to filter out novae independently of any
prior knowledge. This pipeline first selected only objects (defined to
be resolved structures with fluxes significantly higher than the local
median) present in five consecutive observations, to remove rapid
variations. The catalogue was further pruned by selecting against
periodicity, requiring an adequately-sampled primary peak and also
requiring any secondary peak to be an acceptable size. The remaining
candidates were finally required to fit data, rate of decline, colour
and colour-magnitude criteria before being accepted as nova
candidates. \citet{jin} were primarily interested in the cataloguing
of the variable stars in the \PA dataset. They first constructed a
catalogue of variable objects by selecting only (suitably cleaned,
masked, etc.) superpixel \lcs with deviations from their baseline
significant enough in size and duration. Novae were then located by
looking for variable objects matching (within a 3$''$ error-circle)
the positions of novae as published in \textit{IAU Circulars}. Using
these methods, \citet{darn} gave 20 novae and \citet{jin} 12 novae
lightcurves, with 7 novae common to both papers.

\section{An Informal Introduction to Neural Networks}

This section is intended as a brief introduction to the basics of
neural network structure and use as they apply to this paper (for more
details, consult \citet{bish} and \citet{mac}). Neural networks are
pattern-recognition tools composed of neurons (or units) arranged in
layers. Neurons come in three types: input, hidden and output. The
structure of the networks used in this paper is one layer of input
units, one layer of hidden units and one layer of output units. The
neurons in neighbouring layers are fully connected with each other,
and these connections have assigned to them adaptive weights which are
used to calculate the response of a specific neuron to its inputs. The
input data are taken as the values of the input units, and the value
of each hidden unit is then given by the sum over all connections of
the activation value on each input unit, weighted by the weight on the
connection. These activation values are calculated using an activation
function acting on the value of the unit. The values of the output
units are calculated in a similar fashion, except the sum is performed
over all connections between the output unit in question and the
hidden units. In this paper, the activation function is chosen to be
the logistic function, which allows the outputs to be interpreted as
\textit{a posteriori} \probsns.

Before all this can happen, the network must be trained in order to
determine the weights. The weights are initially randomised, and the
network is presented with a training set, made up of sets of input
values (called patterns) for which the desired outputs are known. The
outputs produced by the randomly-weighted net are compared to the
desired values, and the network performance on all patterns is
quantified using an error function, namely the cross-entropy error
\citep{bish, vas1}. A learning function then uses these errors in
conjunction with the values of the hidden units and the
hidden-to-output layer weights in order to update the weights and
hence reduce the output errors. The errors are also propagated back up
to the input-to-hidden layer weights so as to update these weights
with the same goal in mind. This whole process, called
back-propagation, is carried out a number of times (called epochs)
until the desired network performance is reached. With most choices of
learning function, it is possible for the network to become
over-trained on the training set, with the result that performance on
a more general set of inputs is reduced. In this a paper, a special
learning function (see \S \ref{sub:netdet}) is used to avoid this
problem.

The process behind training neural networks is the minimisation of the
error function (as applied to the training set) with respect to the
adaptive weights within the network. This error function may not have
just a global minimum in the multi-dimensional weight-space, but could
have a number of local minima instead or as well. In any case,
networks trained using the exact same training set for the same number
of epochs, but using different initial weights (and therefore
different starting points in this space), will converge to slightly
different final weights. In the case of multiple minima, this means
that networks can follow different error-minimisation paths into
entirely separate minima, some of which might classify the general set
(as opposed to the training set) much better than others. We can turn
this fact to our advantage by using network committees (see
\citet{bish}, \S9.6 and \S10.7), produced by training groups of
networks on the same training set but with initial weights randomly
chosen from a range of values. These networks therefore sample a
region (rather than a point) of the weight-space around the error
function minimum/a, and hence produce a range of results when
classifying the final test set. The results can then be averaged out
over the committee to take account of a whole range of network
`opinions', making sure poor quality networks stuck in high-error
minima don't overly affect the results.

\section{Network Preparation}

\subsection{The Training Set}
\label{sub:TSC}

The ideal training set should contain examples of all forms of stellar
variability we expect the networks to encounter, along with as many
examples of nova \lcs as possible. The usual process is to build the
training set from a comprehensive selection of example nova and
variable star \lcs taken from existing data catalogues. We do not do
this for two reasons. First, there are not enough well-sampled nova
\lcs in the $g$, $r$ and $i$ bands in the standard catalogues for our
purposes. Therefore, we are obliged to simulate nova \lcs from
templates. Second, all of the other forms of variability needed for
the training set are already present in the \PA catalogue, and we can
therefore use the data set itself to provide the non-nova examples
required to build the training set, using a variation on a technique
called \kfold (see below and \citet{bish} \S 9.8.1).

In \kfoldns, the data set is first partitioned into $K$ separate
segments. A network is then trained using a training set containing
all of the data from $K-1$ segments, before being tested on the
remaining segment. This process is then repeated, each time choosing a
different segment to be left out of the training set, until all $K$
choices for the omitted segment have been covered. The test errors are
then averaged out over all $K$ results to create a much more robust
estimate of the network performance, hence providing one of the two
main advantages of using this technique. The second advantage is that
all of the examples in the data set are used in both training and
testing, in effect creating a large training set without the need for
any `external' data. The major disadvantages are that the training
process must be repeated $K$ times, and that some or all of the
training sets will contain nova-type \lcs present in the catalogue
falsely identified as non-nova objects. We therefore use a new
variation on the technique, training networks using just one data
segment before testing the networks on the remaining $K-1$
segments. We believe this is advantageous as it reduces both
processing time and the risk of training set contamination, whilst
still retaining the benefits of normal \kfoldns.

The final form for the training set is 1000 simulated nova light
curves, assigned desired output probabilities of 1, and 1000
randomly-chosen \PA \lcsns, with desired output probability 0. The
decision to use exactly 1000 \PA \lcs is a compromise: 1000 \PA \lcs
should include a sufficient cross-section of the forms of variability
whilst greatly reducing individual training times and keeping the
number of falsely-classified nova examples down to O(1) per training
set.\footnote{There are $\sim$40,000 \lcs in the catalogue, with O(20)
true nova examples present. Hence choosing 1000 \PA examples per
training set gives $\sim$0.5 false nova-type \lcs per set.} The main
drawback to using 1000 \PA \lcs is that the training process must be
repeated $\sim 40$ times, and is therefore quite slow. The number
of nova examples is chosen to overwhelm any falsely-classified novae
and also to create networks biased towards producing false positives
rather than false negatives. Over-representing the novae (as compared
to their natural frequency) in the training set increases the prior
probability of finding a nova in the set, and hence training using
such sets produces networks that are much more likely to misclassify
non-novae as novae than vice-versa (see \S \ref{sub:decbound}). This
is exactly the trend required considering that we are trying to locate
a very rare phenomenon. Of course, the drawback to permitting more
false positives than false negatives is that an additional algorithm
may be needed after the neural network search to root out the
contaminants.

\begin{table}[!t]
\centering
\begin{tabular}{|c|c|c|}
\hline
An et al. ID & Darnley et al. ID & Half-width \\
 & & at 5\% of  \\
 & & Max. Light \\ \hline
25851 & PACN-99-05 & 64.8 \\
26021 & PACN-00-04 & 59.8 \\
26946 & Not present & 72.7 \\
77324 & PACN-01-06 & 99.6 \\
77716 & PACN-00-06 & 45.5 \\
83835 & Not present & 42.0 \\
\hline
\end{tabular}
\caption{\citet{jin} and \citet{darn} identification numbers of the
template novae, along with estimates of decay time.}
\label{templates}
\end{table}

\subsection{Novae Templates}

\begin{figure}[t]
\begin{center}
\includegraphics[height=7cm]{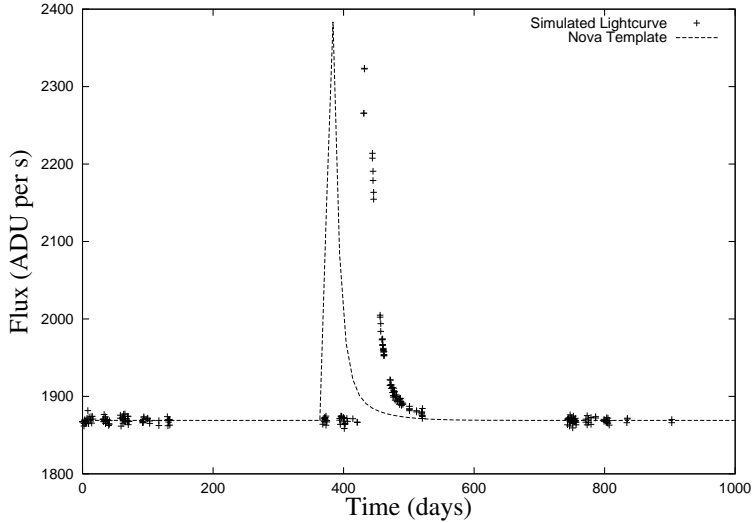}
\caption{Model function for template nova 26021, along with simulated
\lcns.} \label{simLC}
\end{center}
\end{figure}

Six novae identified by \citet{jin} (see Table \ref{templates}) are
chosen as templates. They are selected as having well-sampled peaks
with intermediate decay timescales: their half-widths at 5\% of
maximum light (an indication of the total length of the decay) are all
in the range 40-100 days. Three other novae (An et al. IDs 26277,
78668 and 83479) were also originally included as templates, but their
inclusion reduced the consistency (in terms of both decay timescale
and shape) of the simulated portion of the training set and resulted
in poor final network performance. Note that, due to the limited
timescales covered by the templates and the differences in the \lcs of
CNe of different speed classes, we expect our networks to suffer when
asked to classify novae with much longer or shorter timescales. The
template \lcs are fitted using a model function consisting of a flat
background, a steep linear rise and a function $f(t)$ of the form as
shown below to match the decay:
\begin{equation} \label{eq:sumexp}
f(t)  =  A_1 \exp \left( \frac{-(t - t_{m})}{\tau_1} \right)
+ A_2 \exp \left( \frac{-(t - t_{m})}{\tau_2} \right) + B
\end{equation}
where $A_i$ are the relative sizes of the exponentials,
$t_{m}$ is the time of maximum light, $\tau_i$ are the exponential
decay timescales and $B$ is the value of the background flux. Figure
\ref{simLC} shows an example of such a model function. 

To create the 1000 simulated novae, we repeat the following
procedure. First, a random template is selected, and its peak is
shifted randomly in time within the time limits of the \PA
measurements. A \PA light curve is then chosen at random from the
catalogue, and its sampling times are used to sample the newly-shifted
model function. At this point, we require that there are at least 10
sampling times present in the first 30 days after the peak time of the
shifted model, to ensure that enough of a signal was
present.\footnote{Without this requirement many of the simulated nova
light curves have very small peaks (or none at all). There was
therefore a large constituent group of the training set whose \lcs
were dominated by the random Gaussian fluctuations we added, and so
the networks simply learned to recognise these \lcs instead of the
nova-like \lcs.} A small amount of Gaussian noise is then added to the
sampled, shifted model in order to create simulated novae \lcs\ that
are as similar in form as possible to the original novae (see Figure
\ref{simLC}).

\subsection{Pre-Processing and Network Inputs}
\label{sub:inputs}

The computational power required to use a network grows quickly with
each added input. It is therefore usual to pre-process the
lightcurves, that is, to extract a small number of features from the
data to use as inputs. In this paper, we reduce each \lc to its power
spectrum, before binning and suitably normalising both the individual
\psa \emph{and} the training set as a whole \citep{vas,vas1}.

\begin{figure}[t]
\begin{center}
\includegraphics[height=9cm]{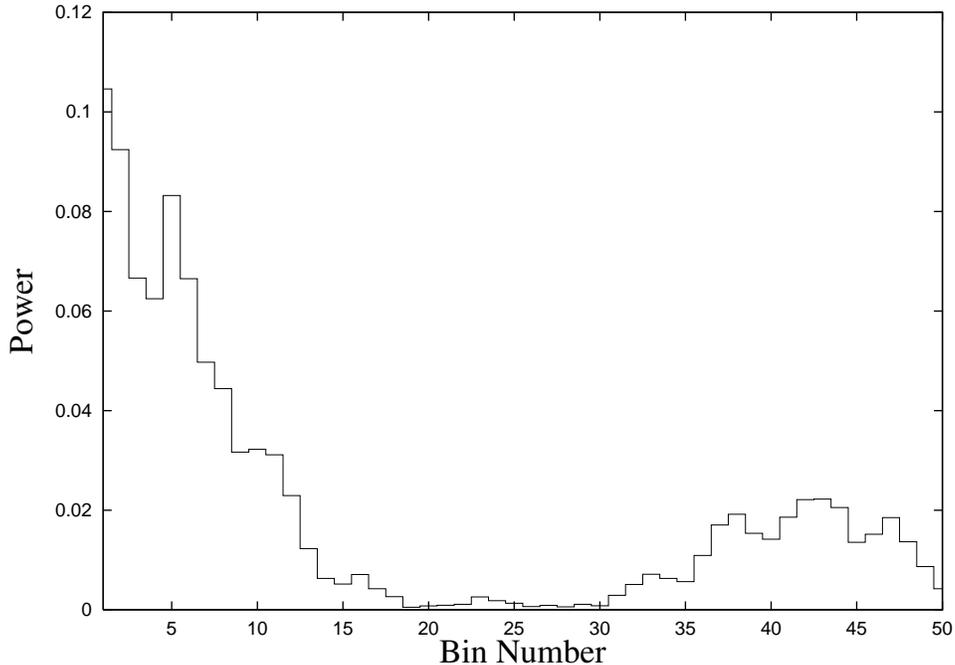}
\caption{Binned, normalised power spectrum (prior to
full-training set normalisation) of a nova \lcns. Higher bin
numbers correspond to higher frequencies.} \label{binnedPS}
\end{center}
\end{figure}

The first reason for reducing the data to their \psa is that the
features that distinguish the nova-type light curves from the other
forms of variability -- i.e., the event timescales, the singular
nature of the eruptions and the shape of the nova peaks -- all
manifest themselves in the \psns. To see this, consider a simplified
nova eruption as a top-hat function of width $w$. The Fourier
transform of a top-hat function of width $w$ in positive
frequency-space is (half) a sinc function, with a central peak of
half-width $\pi/w$. Hence, we expect that the power spectra of our
actual nova eruptions with decay timescales $\tau$ to be distortions
of sinc functions with central peaks of widths of the order of
$\pi/\tau$. From this line of reasoning, we expect the almost singular
nature of the nova eruptions to make their power spectra sinc-like,
with the individual timescales affecting the widths of the sinc peaks,
and the shapes of the outbursts distorting the \psa as a whole. Some
evidence for this can be found in Figure \ref{binnedPS}, which shows
that the nova \psa do indeed resemble sinc functions with roughly
correct peak widths.


A further reason for choosing the \ps is that the features we wish to
select \emph{against}, such as periodicity or random variations,
should also manifest themselves in the \psa of the non-nova objects.
The \ps is also invariant under time-translation of the initial light
curve. Furthermore, the \ps is easily binned, which allows for the
reduction in dimensionality to produce practical networks, although
care \emph{must} be taken to ensure that too much information is not
lost. Due to the uneven time-sampling of the \PA \lcs, we used the
Lomb Periodogram \citep{numrec} to calculate the \psa. The \psa are
determined in the frequency range 0 - 0.3 day$^{-1}$, as this range of
values contains a significant number of CNe \ps features. The \psa are
all binned into 50 constant-width bins, as this results in a
manageable number of network inputs but still retains the resolution
of the original \psans.

The next pre-processing technique is to normalise each individual
binned \psns. This has two positive effects: first, it ensures that
all of the inputs are consistently drawn from within the same range
(from zero to one), and second, it reduces the chances of the networks
classifying two differently-shaped \psa simply because they contain a
similar size peak. Normalising the individual \lcs helps the networks
classify objects on the shapes of their \psans, rather than the size
of any peaks the \psa contain. An example of a binned normalised nova
\ps as it appears at this stage of pre-processing is shown in Figure
\ref{binnedPS}.

The last pre-processing technique is to shift the first input of each
pattern in the training set by the mean of all the first inputs, and
then scale it by dividing by the standard deviation of all of the
first inputs. This is repeated for each input, so that \emph{all} of
the networks' inputs are not only drawn from the same range, but also
have comparable magnitudes, which forces the networks into classifying
the set using \emph{all} of the inputs provided. As an illustrative
example, prior to the introduction of this technique, the nova \ps
typically has low-frequency bin powers a factor of $10^2$ greater than
their high-frequency bin powers (see Figure \ref{binnedPS}). Now we
would consider a 10\% variation in the power in any bin to be equally
important, but a 10\% variation in a high-frequency bin would appear
to the networks to be much less important than a 10\% variation in a
low-frequency bin. By scaling the inputs as described, the networks
classify using the relative, and not absolute, sizes of bin-power
variations between different objects.

\subsection{Network Architecture}
\label{sub:netdet}

The networks used in this paper are all created using the Stuttgart
Neural Network Simulator\footnote{See
http://www-ra.informatik.uni-tuebingen.de/SNNS/}, and are made up of
one layer of 50 input units, one layer of 24 hidden units and one
layer consisting of one output unit (the reasons behind this choice
are given shortly). The units in the hidden layer are fully connected
to both the input and output layers, and the value of the output unit
gives the \textit{a posteriori} \prob that the subject \lc is a nova,
given the weights and the inputs calculated for the subject.  Our
networks use as a learning function resilient back-propagation with
adaptive weight-decay (\rpns). Particularly high adaptive weights
correspond to very strong pattern recognition, and therefore tend to
suggest over-fitting of the training set. During the training process,
\rp therefore automatically allows the highest weights to decay
intelligently so as to keep the network as generally applicable as
possible. Hence, when using \rp, there is no need for the validation
process required by other learning functions (a much fuller
explanation can be found in \cite{bish}, \S 9 and \S 10).

The last choice to make is the number of units. Choosing the number of
input and output units is straightforward: these numbers are simply
determined by the number of inputs (in our case, 50) and outputs (in
our case, one) that the networks receive and produce,
respectively. However, in tasks such as this, it is impossible to
choose the required number of hidden units \nhns\
theoretically. Instead, \nh must be determined experimentally, by
examining the behaviour of the errors produced by networks of
differing \nh in classifying the training set and a new test set (same
form as the training set but totally new \lcsns).

We expect both the training and test errors to decrease at first with
increasing \nhns. Low \nh networks are very simple, so increasing the
number of hidden units increases the network's complexity and hence
ability to map the decision boundary between the classes of
object. However, for some values of \nhns, we expect the behaviour of
the training and test errors to diverge, with the training error
continuing to decrease but the test error either levelling off or
beginning to rise. This differing behaviour occurs because the
networks have become complex enough to start to over-train on the
training set.  The final number of hidden units is therefore chosen to
be the value of \nh at which the training and test error behaviours
diverge, as this gives the best general network performance.

\begin{figure}[t]
\begin{center}
\includegraphics[height=9cm]{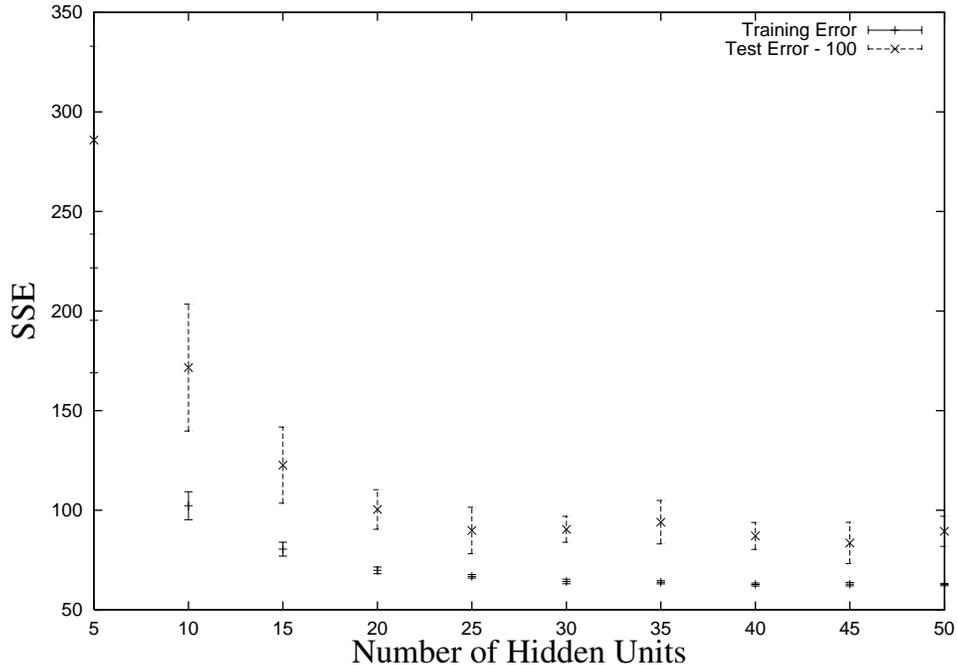}
\caption{The errors on training and test sets
for a range of numbers of hidden units (SSE stands for the sum of the
squared errors of all outputs). NB: test set errors have been shifted
down by 100 to aid comparison.} \label{hiddenunit}
\end{center}
\end{figure}

A plot of the mean errors produced by our networks in classifying the
training and test sets against \nh is shown in Figure
\ref{hiddenunit}. These results are produced by training committees of
ten networks for each value of \nh, with each network given initial
weights drawn randomly from the range -3 to 3. The networks are
trained for 1000 epochs, after which the final errors in classifying
the training set are recorded. The trained networks are then each
tested using the same test set. The training and test errors are
finally averaged out over each committee, thereby providing mean
values to represent more reliably the performance of the different
size networks. The standard deviations are also computed to give some
idea of the the mean error spread.

The first feature to note in Figure \ref{hiddenunit} is that the test
error values are all significantly larger than their corresponding
training errors. This is because there are likely to be numerous \lcs
in the test set of which there are no similar examples in the training
set, due to the random selection of the \PA \lcs included in each
set. This increases the risks of mis-classification. The most
important information to take from the plot is the behaviour of the
errors. For small values of \nh, the behaviour of both the training
and test errors is very similar, as expected. For \nh between 20 and
25, however, the behaviour of the two errors begins to differ: the
training error continues decreasing asymptotically, whereas the test
error levels out within its error bars. We therefore use 24 hidden
units in the networks to produce our final results.

\section{Production of Final Results}

\subsection{The Network Probabilities}
\label{sub:netprobs}

44 committees consisting of 10 networks, each with 50 input units, 24
hidden units and 1 output unit, are created with random initial
weights. These networks are trained using training sets as described
in \S \ref{sub:TSC} for 1000 epochs, taking care to record the \PA
\lcs used and the 50 input means and standard deviations (as described
at the end of \S \ref{sub:inputs}) for each training set. The trained
networks are then used to classify two data sets, which are
pre-processed in the same fashion as the training set but were
normalised using the input means and standard deviations specific to
each committee. The first data set is the cleaned \PA catalogue, and
the second consisted of all of the novae identified by \citet{jin} and
\citet{darn} missing from the catalogue, as listed in Table
\ref{missing}. The initial form of the results is therefore a set of
440 \probs for each \PA object and each previously identified
nova. Each object's results are first averaged out over the ten
networks in each committee, producing 44 committee \probs and errors
for each object, before these values are averaged over the
committees. The \PA objects' \probs and errors are averaged out over
only those committees in whose training sets they did not feature,
whereas the previously-identified novae's values are averaged over all
44 committees.

\begin{table}[t]
\centering
\begin{tabular}{|c|c|}
\hline
Darnley et al. ID & An et al. ID \\ \hline
PACN-99-01$^\dag$ & 10889 \\
PACN-99-02$^\dag$ & 28862 \\
PACN-99-03$^\dag$ & 82483 \\
PACN-99-04$^\dag$ & 93392 \\
PACN-99-07$^\dag$ & 49835 \\
PACN-00-01* & 26946* \\
PACN-01-02* & 83835* \\
\footnotesize{No ID Available} & 26277$^\ddag$ \\
`` & 26285$^\ddag$ \\
`` & 78668$^\ddag$ \\
`` & 79136$^\ddag$ \\
\hline
\end{tabular}
\caption{Identification numbers of the M31 novae, as identified by
Darnley et al. ($^\dag$), An et al. ($^\ddag$) or both *, missing from
cleaned data set.} \label{missing}
\end{table}

\subsection{Decision Boundary Determination}
\label{sub:decbound}

The final task is to set the decision boundary for classification,
that is to determine the \prob value an object must exceed in order to
be classified as a CN. This requires the network's performance to be
quantified in terms of numbers of false positives (\PA objects with
\probs greater than that of the decision boundary) and negatives
(simulated novae with \probs less than that of the decision boundary)
for a range of decision boundary choices. The decision boundary is
chosen so as to optimise the rates at which these false
classifications occur.

\begin{figure}[t]
\begin{center}
\includegraphics[height=9cm]{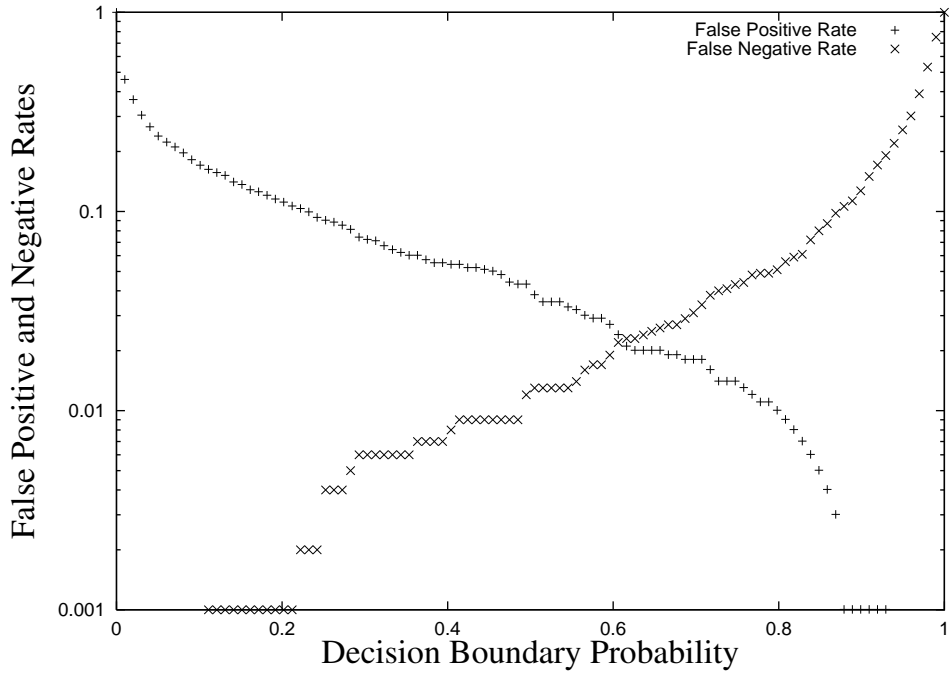}
\caption{Rates of false positive and negative
classifications for a range of decision boundary \prob values.}
\label{falsepn}
\end{center}
\end{figure}

First, a test set is produced using 1000 new simulated novae and the
\PA \lcs used to train one network committee. The test set was then
pre-processed, classified by the other committees, and the results
averaged out in the same way as the \PA catalogue results in \S
\ref{sub:netprobs}. The decision boundary probability $p_{\rm db}$ is
set at different \prob values between 0 and 1, and the numbers of
false positives $N_{\rm fp}(\rm test)$ and negatives $N_{\rm fn}(\rm
test)$ for the test set determined. The results are plotted in Figure
\ref{falsepn}.

If, in our catalogue, the number of non-nova objects $N_{\rm var}^{\rm
tot}(\rm cat)$ were approximately equal to the number of novae $N_{\rm
nov}^{\rm tot}(\rm cat)$, the standard procedure would be to choose
the decision boundary such that the rates of false positives and
negatives are equal. In actuality, however, we expect $N_{\rm
var}^{\rm tot}(\rm cat) \thickapprox 44600$ and $N_{\rm nov}^{\rm
tot}(\rm cat) \thickapprox 20$. This means that if the decision
boundary were chosen to be the point at which the rates $r_{\rm
fp}(\rm test)$ and $r_{\rm fn}(\rm test)$ were equal (i.e. $r_{\rm
fp}(\rm test) = r_{\rm fn}(\rm test) \thickapprox 0.025$), then the
number of expected false positives is $\sim$1100: much bigger than the
number of true novae expected. The choice of decision boundary must
therefore be taken to minimise the number of false positives whilst
ensuring most novae are still detected.  Accordingly, the decision
boundary probability is fixed to be 0.95. At this value, $r_{\rm
fn}(\rm test) \thickapprox 0.2$ from Figure \ref{falsepn}, so we
expect 20 \% of the true novae to be missed. No value for $r_{\rm
fp}(\rm test)$ is available for this $p_{\rm db}$ (probably because
the test set was too small to contain any \PA objects with outputs as
high as 0.95), however an upper bound on the value can be found by
taking the last non-zero value, which is $\sim 0.001$. For this
decision boundary, we therefore expect $< 45$ false positives, a much
more manageable number comparable to the total number of true novae
expected.

\section{The Nova Catalogue}
\label{sub:cat}

The nova catalogue, comprises 47 objects classified by the networks as
having \probs greater than 0.95 of being novae, and is made up of 9
previously identified novae (discussed in \S \ref{sub:oldnovae}), 19
new nova candidates and 19 probable contaminants (all discussed in \S
\ref{sub:newnovae}).

\subsection{Previously Identified Novae}
\label{sub:oldnovae}

\begin{figure}[t]
\begin{center}
\includegraphics[height=10cm]{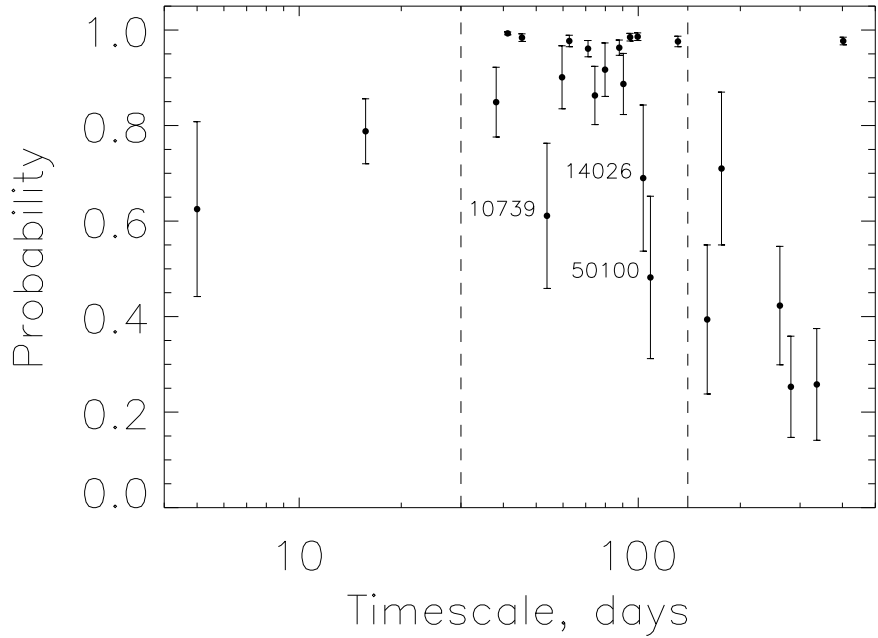}
\caption{The nova probability versus 5\% timescale
for the 25 previously-identified novae. The region within which
highest nova sensitivity is reached is indicated with a dashed line.}
\label{prob_vs_t}
\end{center}
\end{figure}

The average probabilities produced for the 25 CNe previously
identified by An et al. and Darnley et al. are shown in Table
\ref{main_results}. Also included in this table are two decay
timescales: the half-widths of the peaks at $1/e$ and 5\% of maximum
light ($t_e$ and $t_{5\%}$ respectively), chosen to give an indication of
the timescale of the initial ($t_e$) and overall decay ($t_{5\%}$). The
networks trained in this paper correctly identify nine of the novae
(using the criterion from \S \ref{sub:decbound}), with three further
novae falling within their \prob errors' distance of the
classification cut-off. A plot of the \probs assigned to the 25 novae
against their 5\% timescales is shown in Figure \ref{prob_vs_t}.
Examination of this plot indicates two main trends in the data. The
first trend is that the novae that are classified with higher \probs
also have much smaller \prob errors than the mis-classified novae. The
poorly-classified (\probs of 0.7 and lower) novae in particular are
therefore classified much better by some networks than others, which
suggests that their \psa are being confused. The confusion could be
because the \psa of these objects are similar to \PA objects present
in only some network's training sets, or because the networks have
never seen this form of novae before.

\begin{table*}[p]
\centering
\begin{tabular}{|c|c|c|c|c|}
\hline
Object's & Object's & Half-width at & Half-width at & Averaged Network \\
Darnley et al. & An et al. & 1/e of Max. & 5\% of Max. & Response \\
ID & ID & Light (days) & Light (days) & \\ \hline
PACN-99-01$^\dag$ & 10889  & 10.8 & 99.3  & 0.863 $\pm$ 0.061 \\
PACN-99-02$^\dag$ & 28862  & 55.9 & 291.1 & 0.253 $\pm$ 0.106 \\
PACN-99-03$^\dag$ & 82483  & 13.2 & 38.3  & 0.849 $\pm$ 0.073 \\
PACN-99-04$^\dag$ & 93392  & 24.3 & 267.6 & 0.423 $\pm$ 0.124 \\
PACN-99-05*       & 25851* & 7.8  & 64.8  & 0.977 $\pm$ 0.012 $^\bullet$\\
PACN-99-06$^\dag$ & 10739  & 13.2 & 54.0  & 0.611 $\pm$ 0.152 \\
PACN-99-07$^\dag$ & 49835  & 28.6 & 178.5 & 0.710 $\pm$ 0.160 \\
PACN-00-01*       & 26946* & 19.2 & 72.7  & 0.961 $\pm$ 0.017 $^\bullet$\\
PACN-00-02$^\dag$ & 50081  & 72.4 & 433.9 & 0.977 $\pm$ 0.008 $^\bullet$\\
PACN-00-03$^\dag$ & 24225  & 13.6 & 87.8  & 0.963 $\pm$ 0.016 $^\bullet$\\
PACN-00-04*       & 26021* & 26.8 & 59.8  & 0.901 $\pm$ 0.066 $\circ$\\
PACN-00-05$^\dag$ & 50100  & 37.3 & 108.8 & 0.482 $\pm$ 0.170 \\
PACN-00-06*       & 77716* & 10.6 & 45.5  & 0.984 $\pm$ 0.008 $^\bullet$\\
PACN-00-07$^\dag$ & 87092  & 25.3 & 135.5 & 0.976 $\pm$ 0.011 $^\bullet$\\
PACN-01-01*       & 81539* & 79.5 & 159.7 & 0.394 $\pm$ 0.156 \\
PACN-01-02*       & 83835* & 7.7  & 42.0  & 0.993 $\pm$ 0.003 $^\bullet$\\
PACN-01-03$^\dag$ & 14026  & 60.3 & 103.0 & 0.690 $\pm$ 0.153 \\
PACN-01-04$^\dag$ & 82840  & 18.1 & 81.7  & 0.917 $\pm$ 0.056 $\circ$\\
PACN-01-05$^\dag$ & 1881   & 23.9 & 94.7  & 0.985 $\pm$ 0.008 $^\bullet$\\
PACN-01-06*       & 77324* & 27.6 & 99.6  & 0.986 $\pm$ 0.008 $^\bullet$\\
\footnotesize{No ID Available} & 26277$^\ddag$ & 13.5 & 93.3 & 0.887 $\pm$ 0.064 $\circ$\\
``         & 26285$^\ddag$ & 5.0  & 5.1   & 0.625 $\pm$ 0.183 \\
``         & 78668$^\ddag$ & 10.1 & 336.5 & 0.258 $\pm$ 0.117 \\
``         & 79136$^\ddag$ & 0 \footnotesize{(1-point peak)} & 0 & 0.022 $\pm$ 0.007 \\
``         & 83479$^\ddag$ & 3.8 & 16.0 & 0.788 $\pm$ 0.068 \\
\hline
\end{tabular}
\caption{Probability values assigned to novae previously
located by Darnley et al. ($^\dag$), An et al. ($^\ddag$) or both
*. Objects definitely classified as novae by our networks are marked
$^\bullet$. Objects just misclassified (i.e. whose \probs \emph{plus
errors} overlap the decision boundary) are marked $^\circ$.}
\label{main_results}
\end{table*}

\begin{figure}[t]
\begin{center}
\includegraphics[height=20cm]{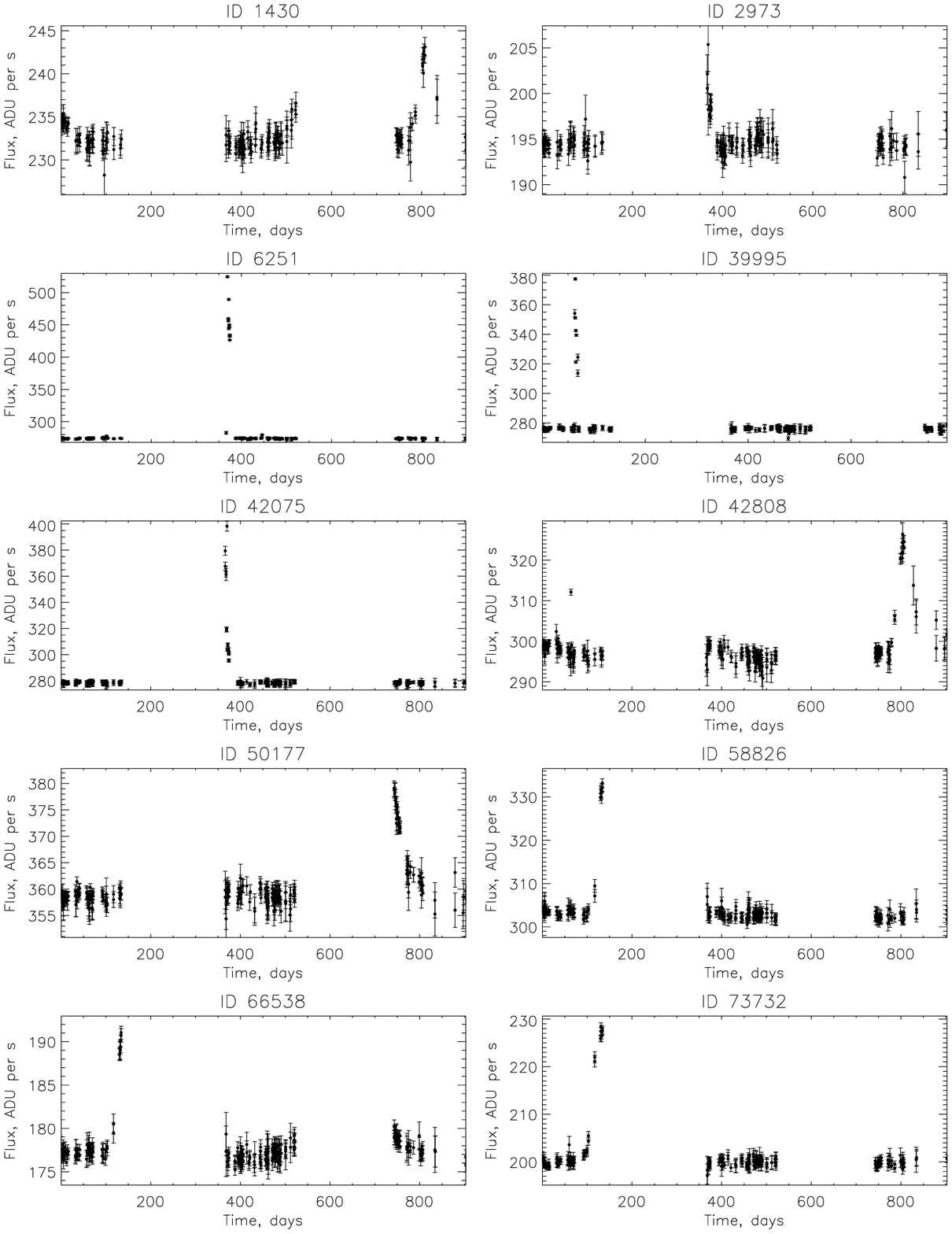}
\caption{Lightcurves of the 19 new nova candidates. The $r$ band flux
in ADU s$^{-1}$ is plotted against time in JD-2451392.5. The 4 strong
candidates are 1430, 42808, 50177 and 74935.}
\end{center}
\end{figure}
\addtocounter{figure}{-1}
\begin{figure}[t]
\begin{center}
\includegraphics[height=20cm]{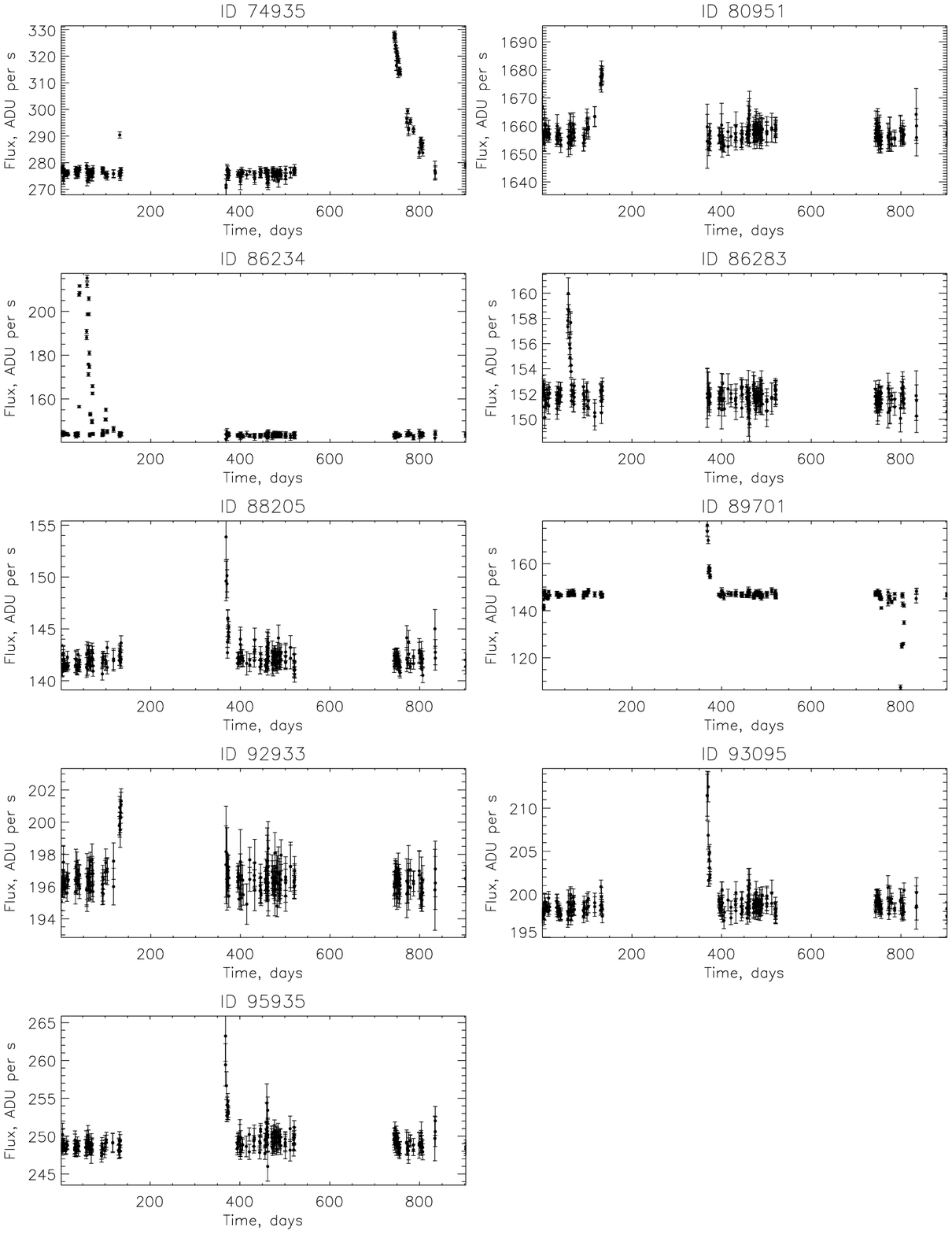}
\caption{(continued). Lightcurves of the 19 new nova candidates. The
$r$ band flux in ADU s$^{-1}$ is plotted against time in
JD-2451392.5. The 4 strong candidates are 1430, 42808, 50177 and 74935.}
\label{new}
\end{center}
\end{figure}

\begin{figure}[t]
\begin{center}
\includegraphics[height=15cm]{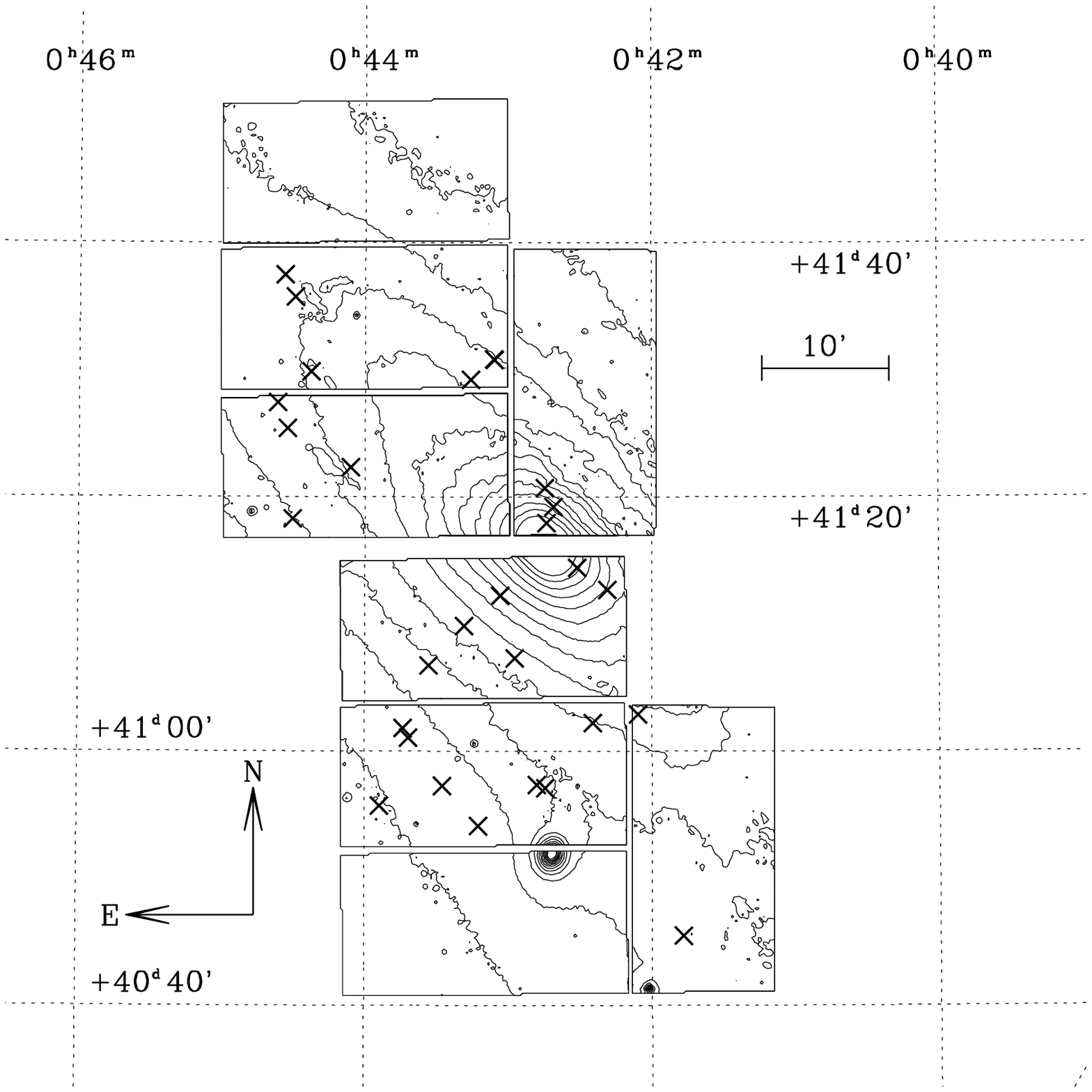}
\caption{Locations of the 19 candidates in Table 4, plus the 9
previously identified novae from Table 3. This is the entire sample of
candidates with a network probability $> 0.95$.}
\label{loc}
\end{center}
\end{figure}

The second trend is that novae with $t_{5\%}$ in the range 30 to 140 days
are generally classified much better than those outside the range,
apart from three exceptions in the range (IDs 10739, 14026 and 50100
(specifically marked in Figure \ref{prob_vs_t})) and one outside (ID
50081). Upon closer inspection of the exceptions within the range,
reasons for their mis-classification become apparent. The \lc of nova
50100 has a very significant second peak and even some evidence for a
third, as well as a confusing bump in the later part of the \lcns. We
therefore do not expect to classify this object well. Lightcurve 14026
actually has a much slower decay than is indicated by its $t_{5\%}$ value
(the reason behind this being its poorly-sampled decay), and so should
be located further right in the plot. Its \lc also features a second
bump in the early stages of its decay. Lightcurve 10739 at first
appears to be ideal for our networks, but its peak is poorly-sampled
near maximum. This seems to hinder the Lomb periodogram, as nova
10739's \ps contains large amounts of high-frequency noise. These
objects therefore should either really not be found in this region of
the plot, or possess features which make them differ from the
template nova \lcs our networks are trained to recognise.

Discarding these objects, the networks correctly identify eight out of
the 13 novae found in the preferred $t_{5\%}$-range ($\sim$62\%
efficiency). Allowing for the error bars on the network outputs, a
further 3 novae fall above the decision boundary ($\sim$ 92 \%
efficiency).  Therefore, the networks can be reliably used to
recognise typical novae with timescales in the range
$30<t_{5\%}\lesssim140$ days, but not outside this range. Note that this
range is actually slightly larger than the range of timescales used in
the template \lcs (i.e. $40<t_{5\%}<100$ days), as the networks can
generalise to some extent. The rapid fall-off of the network's
response for novae with $t_{5\%}$ much greater than 100 days is to be
expected, as slow CNe are much more likely than fast CNe to have decay
fluctuations and secondary peaks and hence be significantly different
to the template novae. The low-$t_{5\%}$ fall-off of the network's
response is also expected, as it corresponds to the \ps range becoming
too small to fit in the main features of the novae's sinc-like \ps
(see \S \ref{sub:inputs}). These fall-offs mean that in order to
recognise novae with $t_{5\%}$ values outside of the preferred range, we
will have to alter the pre-processing techniques.

Additionally, one further nova outside the preferred timescale range
is detected. We note that the positive classification of \lc 50081 is
highly inconsistent with the results for other slow novae. Its \lc is
well-sampled, and clearly belongs to a very slow nova, and yet its \ps
appears to be recognisable to the neural networks. We currently have
no explanation as to why the networks should pick it up, as nothing
comparable to it appears in the training set, but as it appears in a
region where little response is expected, it is more of an added bonus
than a troubling anomaly~\footnote{A duplicate of 50081, namely 50153,
is also detected. However, it is removed from the list of new nova
candidates by human intervention.}

\begin{table*}[p]
\centering
\begin{tabular}{|c|c|c|c|c|c|c|}
\hline
Object & RA & dec &Half-width at & Half-width at & Network & Nova\\
ID  & (hr:min:sec)& (deg:min:sec) & 1/e of Max. & 5\% of Max. &
Probability & Features \\
 & & & Light (days) & Light (days) & & \\ \hline
1430 &  00:44:36.564 & 41:27:24.159 & 47.7 & 91.6 &  0.966 $\pm$ 0.016 & M  \\
2973 &  00:44:30.446 & 41:18:13.510 & 9.6 & $\sim$25 & 0.959 $\pm$
0.041 & D, F \\
6251 &  00:44:05.978& 41:22:19.384  & 11.0 & 17.0 & 0.975 $\pm$ 0.012
& P, F\\
39995 & 00:44:33.854 & 41:37:27.775 & 5.0 & 5.8 & 0.971 $\pm$ 0.018 &
P, F \\
42075 & 00:44:29.306 & 41:35:42.513 & 1.5 & 15.1 & 0.975 $\pm$ 0.012 &
P, F \\
42808 & 00:44:22.572 & 41:29:50.579 & 31.9 & 89.5 & 0.955 $\pm$ 0.020 & M \\
50177 & 00:43:15.816 & 41:29:12.045 & 29.3 & 64.8 & 0.955 $\pm$ 0.011 & M \\
58826 & 00:42:05.645 & 41:02:49.409 & \footnotesize{No decay} & \footnotesize{No decay} & 0.986 $\pm$ 0.005 & R \\
66538  & 00:41:46.976 & 40:45:28.867& `` & `` & 0.984 $\pm$ 0.005 & R \\
73732 & 00:43:33.160 & 41:06:44.146 & `` & `` & 0.954 $\pm$ 0.020 & R \\
74935 & 00:43:18.538 & 41:09:48.496 & 36.7 & 84.4 & 0.954 $\pm$ 0.009 & M \\
80951 & 00:42:31.148 & 41:14:25.462 & \footnotesize{No decay} &
\footnotesize{No decay} & 0.961 $\pm$ 0.017 & R \\
86234 &  00:43:41.720 &41:01:04.352 & `` & `` & 0.989 $\pm$ 0.014 & P,
F\\
86283 &00:43:44.186 & 41:01:48.806& 6.0 & 6.0 & 0.968 $\pm$ 0.016 & D, F
\\
88205 &00:43:27.490 &40:57:16.057 & $\sim$3 & $\sim$10 & 0.980 $\pm$
0.009 & D, F \\
89701 & 00:43:12.483 & 40:54:05.979 & 3.0 & 107.8 & 0.953 $\pm$ 0.026
& D, F \\
92933 & 00:42:48.081 & 40:57:20.327 &
\footnotesize{No decay} & \footnotesize{No decay} & 0.954 $\pm$ 0.023 & R \\
93095 & 00:42:44.604 & 40:57:04.521 & 3.6 & 52.1 & 0.979 $\pm$ 0.008 &
D, F \\
95935 &  00:42:24.705& 41:02:12.578 & $\sim$3 & $\sim$10 & 0.977 $\pm$
0.015 & D, F \\
\hline
\end{tabular}
\caption{The IDs, right asencion and declination (J2000.0), network
probabilities, decay timescales (where possible) and nova features (M
= most or all, R = rise only, D = decay only and P = peak only) of the
19 new nova candidates. F = possible fake, as judged by examination of
the image frames.} \label{newnovtab}
\end{table*}

\subsection{New Nova Candidates}
\label{sub:newnovae}

The nova catalogue also contains 19 \lcs which, upon inspection, are
either recognisable as novae or exhibit some nova characteristics, and
hence can be classified as candidates for newly discovered novae. The
IDs, location and \probs of these 19 candidates are listed in
Table~\ref{newnovtab}, while their \lcs are displayed in
Figure~\ref{new}.  The candidates can be roughly separated into four
groups according to which nova features they exhibit. The first group
consists of 1430, 42808, 50177 and 74935. Their lightcurves contain
most or all of the desired features, and hence make excellent nova
candidates.

The second group have \lcs where only the first few measurements of a
rise towards a peak are present (IDs 58826, 66538, 73732, 80951 and
92933), with no sampling of the decay. The third group have \lcs with
samples present which suggest some form of decay from a peak, but no
measurements of the rise or peak itself (IDs 2973, 86283, 88205,
89701, 93095 and 95935). The fourth group have \lcs which feature
prominent, sharp peaks but not much clear evidence for the
characteristic nova rise or decay (IDs 6251, 39995, 42075 and 86234),
and which could therefore be very fast novae or simply instrumental
defects. It is difficult to say for certain that objects in these
three groups are novae without more data. The locations of the 19
candidates in Table~\ref{newnovtab}, together with the 9 candidates in
Table~\ref{main_results} are shown in Figure~\ref{loc}, superposed on
the optical isophotes of M31. These are all the candidates with a
network probability $> 0.95$.

The difference images of all 19 candidates have been examined and the
PSFs constructed. If the PSF is not roundish with a size controlled by
the seeing, then this suggests that the candidates may be
fakes. Performing this test yields the result that perhaps 10 of the
candidates are spurious (2973, 6251, 39995, 42075, 86234, 86283,
88205, 89701, 93095 and 95935).

Finally, the nova catalogue also contains 19 contaminants that appear
to be true variable objects, and are primarily made up of the \lcs of
superpixels covering periodic stars such as Miras and Cepheids,
although many \lcs exhibit some other superposed form of variability.


\section{Conclusions}
\label{section:conc}

This paper has presented working neural networks for the
identification of fast CNe. The use of \kfold and the choice of
pre-processing techniques (i.e. reducing the \lc to a suitably binned
and normalised \psns) \emph{has} produced a set of neural networks
capable of detecting the fast classical nova present in the
POINT-AGAPE survey. This conclusion is borne out by the consistently
high nova \probs assigned to the previously-identified novae with
$30<t_{5\%}\lesssim140$ days, the detection of 4 strong new nova
candidates in the \PA catalogue and a further 15 possible
candidates. This adds further weight to the claims by a number of
authors \citep{wozniak,vas,vas1,brett} that neural networks offer a
promising solution to the problem of lightcurve identification in
massive variability surveys.

The variation of \kfold used in this paper is new and particularly
well-adapted to the search for rare objects in a large dataset.
Usually, in \kfoldns, the data set is first partitioned into $K$
separate sets. A network is then trained using a training set
containing all of the data from $K-1$ segments, and tested on the
remaining data.  Our variation on this technique is to train the
networks using just one \PA data segment before testing the networks
on the remaining $K-1$ segments. This is beneficial as the processing
times is substantially reduced. In many circumstances, there would be
a risk of training set contamination using this variation on
\kfoldns. However, CNe are very scarce in the \PA dataset.  So, the
\PA \lcs themselves can be used for the non-nova examples in the
training set with little risk of contamination. The nova examples are
produced in the training set must be produced with templates.  This
method therefore can be used to find any rare lightcurves in a massive
variability survey, provided suitable templates exist.

Nonetheless, the networks cannot be used in their current form to
obtain a nova-rate for M31. Very fast novae are missing because the
the \PA sampling rate is just not good enough to detect them.  As
demonstrated by Figure \ref{prob_vs_t}, the networks also do not
detect enough slow, bumpy novae. Furthermore, these novae are more
often than not assigned high \probsns, yet these \probs fall below the
classification cutoff because the networks produce too many false
positives.  The difficulty here is that artifical templates for slow
novae are harder to construct, as they exhibit a greater morphology in
the declining part of the curve. The best way to overcome this is to
use known examples of slow CNe as part of the training
set. Unfortunately, there are very few such lightcurves available in
the $g$, $r$ and $i$ passbands of the \PA survey. This however may
become possible in the future using transformed colours.  The
extension of the networks to slow novae may also require modifications
to the pre-processing technique, as the \psa of slow novae are
different (less sinc-like) to those of fast novae.

Finally, it is worth mentioning the limiting factor for detection of
fast nova is actually the temporal sampling of the POINT-AGAPE
dataset. As fast CN are the brightest CN, they are still easy to
detect even against the bright bulge of M31. Although we have not
carried out a full efficiency analysis, it is clear that the networks
successfully detect the CN types on which the system was trained, up to
the limit imposed by the temporal sampling.

\acknowledgements 
VB and EJK are supported by the Particle Physics and Astronomy
Research Council of the United Kingdom, while JA is supported by the
Leverhulme Trust. Work by AG is supported by NSFgrant 02-01266.  We
thank all the members of the POINT-AGAPE collaboration for access to
their data.

\clearpage

\end{document}